\documentclass[twocolumn, aps, prb]{revtex4}
\usepackage{graphicx}
\usepackage{amssymb}
\usepackage{amsmath}
\def\Journal #1,#2,#3,#4#5#6#7{#1 {\bf #2}, #3 (#4#5#6#7)}

\def\Vec#1{{\bf #1}}
\def\GVec#1{\mbox{\boldmath $#1$}}
\def\vare{\varepsilon}

\def\partd#1#2{\frac{\partial #1}{\partial #2}}

\begin{document}
%
\title{Chiral orbital current and anomalous magnetic moment
in gapped graphene}
\author{Mikito Koshino}
\affiliation{
Department of Physics, Tohoku University, 
Sendai, 980--8578, Japan}
\date{\today}
%
\begin{abstract}
We present a low-energy effective-mass theory to describe
chiral orbital current and anomalous magnetic moment
in graphenes with band gap and related materials.
We explicitly derive a quantum mechanical current distribution
in general Bloch electron systems,
which describes a chiral current circulation supporting
the magnetic moment.
We apply the formulation to gapped graphene monolayer, bilayer 
and ABC-stacked multilayers,
to show that the chiral current is opposite between different valleys,
and corresponding magnetic moment
accounts for valley splitting of Landau levels.
In gapped bilayer and ABC multilayer graphenes, in particular,
the valley-dependent magnetic moment is responsible for
huge paramagnetic susceptibility at low energy, which enables
a full valley polarization up to relatively high electron density.
The formulation also applies to the gapped surface states
of three-dimensional topological insulator,
where the anomalous current is related to
the magneto-electric response in spatially-modulated potential.
\end{abstract}
%
\maketitle
%
\section{Introduction}
\label{sec_intr}
%

The magnetic moment in an electronic system consists of two distinct 
factors due to spin and orbital motion of electrons.
In solids, the spin magnetic moment is enhanced by          
anomalous factor caused by to the orbital effect,
resulting in increase of $g$-factor. \cite{Yafet_1963,Kittel_1963}
Graphene \cite{Novoselov_et_al_2004a,Novoselov_et_al_2005a,Zhang_et_al_2005a}
has an intriguing counterpart of spin, which is 
associated with valley pseudo-spins,
i.e., degree of freedom corresponding to different points
in the Brillouin zone called $K_+$ and $K_-$ valleys.
Specifically, when the band gap is opened by an asymmetric potential
breaking the sublattice symmetry,
the graphene electrons have 
anomalous magnetic moment opposite in different valleys
similarly to real spin.
\cite{Xiao_et_al_2007,Yao_et_al_2008,Koshino_and_Ando_2010}
Generally the anomalous magnetic moment
is closely related to the geometric nature of the Bloch band,
and has been argued in relation to Berry phase.
\cite{Chang_and_Niu_1996,Thonhauser_et_al_2005,Ceresoli_et_al_2006,Xiao_et_al_2005,Xiao_et_al_2010}
Previously we calculated the orbital susceptibility
in gapped monolayer and bilayer graphenes, and showed that
the susceptibility near $K_\pm$ point,
where the dispersion is quadratic,
is contributed from the Pauli paramagnetism caused by the valley pseudo-spin.
\cite{Koshino_and_Ando_2010}

In this paper, to understand the physical origin of pseudo-spin magnetic moment,
and also to investigate the pseudo-spin magnetic moment
in various electronic structures other than quadratic dispersion,
we develop a general low-energy effective-mass theory to describe
anomalous current density supporting the magnetic moment.
We explicitly derive a quantum mechanical current distribution
in general Bloch electron systems,
which describes chiral current circulation
for each eigenstate. 
Using the formula, we actually calculate the valley-dependent chiral current 
in gapped graphene monolayer, 
bilayer \cite{Novoselov_et_al_2006a,Ohta_et_al_2006a,Castro_et_al_2007b,Oostinga_et_al_2008a} 
and ABC-stacked multilayers \cite{Lipson_et_al_1942,Haering_1958,McClure_1969}.
The valley-dependent magnetic moment 
exactly gives the valley splitting of Landau levels,
generalizing our previous results limited to 
the quadratic dispersion. \cite{Koshino_and_Ando_2010}
In gapped bilayer \cite{Koshino_and_Ando_2010} 
and ABC multilayers, in particular,
the valley splitting and diverging density of states at the band bottom 
result in a huge paramagnetic susceptibility,
enabling  a full valley polarization 
up to relatively high electron density 
of the order of $10^{12}$ cm$^{-2}$
at a magnetic field of $\sim$ 1 T.

The formulation also allows to include the external potential
field within the low-energy approximation,
and thus useful to investigate the chiral current in disordered systems 
and also finite systems bound by potential barrier.
It also applies to the gapped surface states
of the three-dimensional topological insulator,
where the anomalous current describes
the magneto-electric response in a spatially-modulated potential.
\cite{Qi_et_al_2008,Essin_et_al_2009,Hasan_and_Kane_2010,Qi_and_Zhang_2010}

Paper is organized as follows.
In Sec.\ \ref{sec_chiral}, we present 
the general effective mass description of
the anomalous current density for Bloch electrons.
We apply this to asymmetric monolayer, bilayer and
ABC multilayer graphenes
in Sec.\ \ref{sec_mono}, Sec.\ \ref{sec_bi} and Sec.\ \ref{sec_abc}, 
respectively, to describe the chiral current circulation, 
magnetic moment and valley splitting of Landau levels.
In Sec.\ \ref{sec_pseudo}, we calculate
the magnetic susceptibility and argue the role of the 
anomalous magnetic moment.
We describe in Sec.\ \ref{sec_hall} 
the current distribution in spatially modulated external potential,
and formulate it in terms of a response function 
analogous to the Hall conductivity.
The conclusion is given in Sec.\ \ref{sec_concl}.



%
\section{Anomalous orbital current}
\label{sec_chiral}
%

We consider 
a Bloch electron system described by
an effective-mass Hamiltonian matrix ${\cal H}_{mm'}(\Vec{p})$,
where $\Vec{p}$ is the crystal momentum, and $m$ and $m'$ are band indeces. 
We assume that the Hamiltonian is diagonalized at $\Vec{p}=0$ as
\begin{eqnarray}
{\cal H}_{mm'}(0) = \vare_{m}^{0} \delta_{mm'},
\end{eqnarray}
and, for simplicity, that there are no degeneracy at $\Vec{p}=0$.
In presense of the external potential $V(\Vec{r})$,
the effective-mass wavefunction $\Vec{F}(\Vec{r})$
obeys the Schr\"{o}dinger equation
\begin{eqnarray}
\sum_{m'} {\cal H}_{mm'}({\Vec{p}}) F_{m'}(\Vec{r})
= [\varepsilon - V(\Vec{r})] F_{m}(\Vec{r}),
\label{eq_schro}
\end{eqnarray}
where ${\Vec{p}} = -i \hbar \nabla$, 
and $\vare$ is the eigen energy.
We assume $|V|\ll|\vare^{0}_m - \vare^{0}_{m'}|$,
so that the states of different bands are not strongly mixed.

We focus on an eigenstate near $\vare = \vare^{0}_{n}$
of the particular band $n$.
Then the wavefunction mainly has its amplitude on $F_n$.
By the first-order perturbation, the amplitude at $F_{m \neq n}$
can be written in terms of $F_n$ as
\begin{eqnarray}
 F_{m}(\Vec{r}) \approx
\frac{{\cal H}_{mn}({\Vec{p}})}{\vare^{0}_{n} - \vare^{0}_{m}}
 F_{n}(\Vec{r}).
\label{eq_Fm}
\end{eqnarray}
The Schr\"{o}dinger equation, Eq.\ (\ref{eq_schro}), then becomes
\begin{eqnarray}
&& [ {\cal H}^{\rm (eff)}_{n}({\Vec{p}}) + V(\Vec{r})] F_{n}(\Vec{r}) 
= \varepsilon  F_{n}(\Vec{r}),
\end{eqnarray}
with the effective Hamiltonian 
\begin{eqnarray}
&& {\cal H}^{\rm (eff)}_{n}({\Vec{p}}) = 
{\cal H}_{nn}({\Vec{p}}) + 
\sum_{m\neq n}
\frac{{\cal H}_{nm}({\Vec{p}}){\cal H}_{mn}({\Vec{p}})}
{\vare^{0}_{n} - \vare^{0}_{m}}.
\label{eq_H_red}
\end{eqnarray}
Correspondingly, we can define the effective velocity operator 
\begin{eqnarray}
{v}^{\mu{\rm(eff)}}_{n}  &= &
\partd{{\cal H}^{\rm (eff)}_n({\Vec{p}})}{{p}_\mu},
\end{eqnarray}
and the local current density operator
\begin{eqnarray}
{j}_n^{\mu{\rm(eff)}} (\Vec{R}) &=&
-\frac{e}{2} 
\left\{{v}^{\mu{\rm(eff)}}_n, \delta(\Vec{r}-\Vec{R})\right\}
\nonumber\\
&& \hspace{-15mm}
=
-\frac{e}{2}
\Biggl[
\left\{
v^\mu_{nn},
\delta(\Vec{r}-\Vec{R})
\right\}
+
\sum_{m\neq n}
\frac{1}{\vare^{0}_{n} - \vare^{0}_{m}}\times
\nonumber\\
&& \hspace{-3mm}
\left\{
(v^\mu_{nm}{\cal H}_{mn} + {\cal H}_{nm}v^\mu_{mn}),\,
\delta(\Vec{r}-\Vec{R})
\right\}
\Biggr],
\label{eq_jred}
\end{eqnarray}
where $\{a,b\} = ab + ba$ is the anti-commutator, and
\begin{eqnarray}
 {v}^\mu_{mm'}  =
\partd{{\cal H}_{mm'}({\Vec{p}})}{{p}_\mu}.
\label{eq_v}
\end{eqnarray}

${j}_n^{\mu{\rm(eff)}}$
actually covers only a part of the total current density
even in the low-energy limit.
The original current density operator is given by
\begin{eqnarray}
{j}^\mu (\Vec{R}) =
-\frac{e}{2} 
\left\{{v}^{\mu}, \delta(\Vec{r}-\Vec{R})\right\},
\end{eqnarray}
where $v^\mu$ is a matrix defined by Eq.\ (\ref{eq_v}).
The expectation value of $j^\mu$
for a given state $\Vec{F}$ near $\vare^{0}_n$ is written as
\begin{eqnarray}
\langle {j}^\mu (\Vec{R}) \rangle
&=&
\sum_{mm'} 
\int d\Vec{r} F_m^*(\Vec{r}) [{j}^\mu (\Vec{R})]_{mm'}
F_{m'}(\Vec{r})
\nonumber\\
&\approx&
\int d\Vec{r} F_n^*(\Vec{r}) 
{j}^\mu_n (\Vec{R})
F_{n}(\Vec{r}).
\label{eq_jav}
\end{eqnarray}
In the second equation we used Eq.\ (\ref{eq_Fm}),
and defined,
\begin{eqnarray}
&&{j}^\mu_n (\Vec{R})
=
-\frac{e}{2}
\Biggl[
\left\{
v^\mu_{nn},
\delta(\Vec{r}-\Vec{R})
\right\}
+
\sum_{m\neq n}
\frac{1}{\vare^{0}_{n} - \vare^{0}_{m}}\times
\nonumber\\
&& \quad
\Bigl(
\left\{
v^\mu_{nm},
\delta(\Vec{r}-\Vec{R})
\right\}
{\cal H}_{mn}
+
{\cal H}_{nm}
\left\{v^\mu_{mn},
\delta(\Vec{r}-\Vec{R})
\right\}
\Bigr)
\Biggr]
\nonumber\\
&&=
 {j}^{\mu {\rm (eff)}}_n (\Vec{R})
\nonumber\\
&& \quad 
-\frac{e}{2}
\sum_{m\neq n}
\frac{1}{\vare^{0}_{n} - \vare^{0}_{m}}
\left(
v^\mu_{nm}
[\delta(\Vec{r}-\Vec{R})
,{\cal H}_{mn}] + {\rm h.c.}\right).
\nonumber\\
\label{eq_j}
\end{eqnarray}
${j}^\mu_n$ is not equivalent with $ {j}^{\mu {\rm (eff)}}_n$
since ${\cal H}_{mn}$ and $\delta(\Vec{r}-\Vec{R})$
do not generally commute. As shown in the following,
the second term, called anomalous current
in the following, is responsible for the chiral current circulation
in gapped graphenes.

The similar argument is available for the orbital magnetic moment.
The operator of the magnetic moment perpendicular to the layer is defined as
\begin{equation}
 m = -\frac{e}{2c}(x v^y - y v^x).
\end{equation}
Similarly to Eq.\ (\ref{eq_jav}),
the expectation value of $m$
for a state of the band $n$ can be written as
\begin{eqnarray}
\langle m \rangle
&\approx&
\int d\Vec{r} F_n^*(\Vec{r}) 
m_n F_{n}(\Vec{r}), 
\end{eqnarray}
where 
$m_n$ is the effective magnetic moment,
\begin{eqnarray}
m_n
&=&
-\frac{e}{2c}
(
x v_n^{y{\rm (eff)}}
-y v_n^{x{\rm (eff)}}
)
\nonumber\\
&& -\frac{e\hbar}{2 c}
\sum_{m\neq n}
\frac{1}{i}
\frac{v^x_{nm}v^y_{mn} - v^y_{nm}v^x_{mn}}
{\vare^{0}_{n} - \vare^{0}_{m}}
.
\nonumber\\
\label{eq_m}
\end{eqnarray}
The first term is the magnetic moment given by
the orbital current $j^{\mu{\rm (eff)}}_n$.
The second term is the extra magnetic moment coming from the anomalous current,
and coincides with the expression of magnetic moment
which enhances the $g$-factor
in a conventional semiconductor physics.
\cite{Yafet_1963,Kittel_1963}

While we include a diagonal scalar potential $V(\Vec{r})$
in above argument, an off-diagonal potential is generally possible
in systems such as graphene with a random vector potential.
As long as the potential term enters the Hamiltonian 
in a form of ${\cal H}_{mn} + V_{mn}(\Vec{r})$, as 
in random vector potential for graphene,
the expression of the chiral current Eq.\ (\ref{eq_j})
is not influenced since $V_{mn}$ 
commutes with $\delta(\Vec{r}-\Vec{R})$,
and also does not alter the velocity operator $v^\mu_{mn}$.

%
\section{Monolayer graphene}
\label{sec_mono}
%

Graphene is composed of a honeycomb network of carbon atoms, where a
unit cell contains a pair of sublattices, denoted by $A$ and $B$.
Low-energy electronic states are described by the effective Hamiltonian,
\cite{McClure_1956a,Slonczewski_and_Weiss_1958a,DiVincenzo_and_Mele_1984a,Semenoff_1984a,Ando_2005a,Shon_and_Ando_1998a,Zheng_and_Ando_2002a,Gusynin_and_Sharapov_2005a,Peres_et_al_2006a}
\begin{equation}
{\mathcal H}({\Vec{p}}) = 
\begin{pmatrix} 
\Delta & v p_-
\\ v p_+  & -\Delta 
\end{pmatrix} ,
\label{eq_H_mono}
\end{equation}
where ${p}_\pm = \xi{p}_x \pm i{p}_y$,
$\xi = \pm$ is the valley index
corresponding to $K_\xi$ point in the Brillouin zone,
and $\Vec{p}$ is the momentum measured from the $K_\xi$.
The matrix works on two-component 
envelope wave function $(F_A(\Vec{r}), F_B(\Vec{r}))$
at the $A$ and $B$ sublattices, respectively.
The diagonal terms $\pm\Delta$, opening the energy gap at Dirac point,
is given by the potential asymmetry between $A$ and $B$ sites, 
which can arise in a certain substrate material for instance. 
\cite{Zhou_et_al_2007a,Zhou_et_al_2008b}
The band velocity is $v \approx 1\times 10^6$ m/s. 

The surface states
of the three-dimensional topological insulator of Bi$_2$Se$_3$
family is also described by a similar Hamiltonian
to Eq.\ (\ref{eq_H_mono}), where $({p}_x,{p}_y)$
is rotated to $({p}_y,-{p}_x)$. \cite{Hasan_and_Kane_2010,Qi_and_Zhang_2010} 
The rotation of vector ${\Vec{p}}$ 
is compensated by the spinor rotation
and does not affect the following argument.
There is only single valley index,
and the diagonal term $\Delta$ appears
only when the time-reversal symmetry is broken,
for instance, by attaching a ferromagnetic material. 
\cite{Qi_et_al_2008,Essin_et_al_2009} 

We assume $\Delta > 0$ and consider a state near the electron band bottom 
$\vare=\Delta$. The wave amplitude is then mainly 
concentrated on the first component $F \equiv F_A$.
The reduced Hamiltonian for $F$ becomes apart from the constant energy,
\begin{equation}
 {\cal H}^{\rm (eff)}({\Vec{p}}) = \frac{{p}^2}{2m^*},
\label{eq_H_mono_red}
\end{equation}
with the effective mass,
\begin{eqnarray}
  m^* = \frac{\Delta}{v^2}.
\label{eq_eff_mass}
\end{eqnarray}
Applying Eq.\ (\ref{eq_j}), the local current density is written as
\begin{eqnarray}
\langle \Vec{j}(\Vec{r}) \rangle
&=&  
-\frac{e\hbar}{m^*}\,
{\rm Im}(F^* \nabla F)
- \xi \frac{e\hbar}{2m^*}
 (-\Vec{e}_z \times \nabla) |F|^2
\nonumber\\
\label{eq_j_mono}
\end{eqnarray}
where
 $\nabla = (\partial/\partial x, \partial/\partial y, 0)$,
and $\Vec{e}_z = (0,0,1)$.
The first term 
is the usual current density, 
corresponding to $j^{\rm (eff)}$
of Eq. (\ref{eq_j}).
The second term is the anomalous component, and denoted as $\Vec{j}_c$
in the following.
It flows perpendicularly to the gradient of the density $|F|^2$,
and thus it circulates on a closed loop and does not 
contribute to the electron transport.
The direction is opposite between $\xi = \pm$.
It is written in terms of equivalent local
magnetic moment $\GVec{\mu}$ as
\begin{eqnarray}
\langle \Vec{j}_c(\Vec{r}) \rangle
&=&  
  c\, \nabla \times \GVec{\mu}(\Vec{r}),
\nonumber\\
\GVec{\mu}(\Vec{r}) &=& - \xi \frac{e\hbar}{2m^*c}|F|^2 \Vec{e}_z.
\label{eq_mr_mono}
\end{eqnarray}
For the valence band electron, 
a similar calculation shows that 
the first term of Eq.\ (\ref{eq_j_mono})
flips the sign while the second term remains unchanged.

The expression of the magnetic moment operator, 
Eq.\ (\ref{eq_m}), becomes
\begin{eqnarray}
 && m 
=
-\frac{e}{2m^*c}(x{p}_y - y{p}_x) 
-\xi\frac{e\hbar}{2m^*c},
\label{eq_m_mono}
\end{eqnarray}
where the first and second terms corresponds to those 
of Eq. (\ref{eq_j_mono}), respectively.
The second term, now denoted as $m_c$,
is the magnetic moment induced by the anomalous current
and coincides with the integral of $\GVec{\mu}(\Vec{r})$
of Eq.\ (\ref{eq_mr_mono}) over the space.
It should be noted that $m_c$ is constant
regardless of the detail of the wavefunction.
This is analog of spin magnetic moment of bare electron system 
with $\xi$ being the spin index, 
while in graphene
this is mimicked by the valley-dependent chiral orbital current. 
The expression agrees with an intrinsic magnetic moment
in the semi-classical picture,  that attributed to
the self-rotation of the wave packet. \cite{Xiao_et_al_2007} 

The valley pseudo-spin magnetic moment $m_c$
produces the pseudo-spin Zeeman energy in presence of a 
magnetic field, 
and this accounts for the valley splitting of Landau levels 
in graphene. \cite{Koshino_and_Ando_2010}
This can be checked by
considering the Hamiltonian in a uniform external field $\Vec{B}$,
or ${\cal H}({\GVec{\pi}})$ in Eq.\ (\ref{eq_H_mono}),
where ${\GVec{\pi}}\equiv {\Vec{p}}+ e\Vec{A}/c$
with the vector potential $\Vec{A}$ giving 
$\Vec{B} = \nabla \times \Vec{A}$.
Noting the relation $[\pi_x,\pi_y] =-i\hbar e B/c$,
the reduced Hamiltonian for the A site 
near $\varepsilon = \Delta$ is written as \cite{Koshino_and_Ando_2010}
\begin{eqnarray}
\mathcal{H}^{\rm (eff)}({\GVec{\pi}}) 
&\approx& 
\frac{v^2}{2\Delta} \pi_- \pi_+
=\hbar \omega_c \left(\hat{n} + \frac{1}{2} + \frac{\xi}{2}\right),
\label{eq_H_mono_red_B}
\end{eqnarray}
where $\omega_c = eB/(m^*c)$,
$\pi_\pm = \xi\pi_x \pm i \pi_y$, $\hat{n} = a^\dagger a$,
$a = (2\hbar e B/c)^{-1/2} ({\pi}_x - i{\pi}_y)$ is 
the annihilation operator of Landau level,
and we used the relation ${\pi}^2 = (2\hbar eB/c)(\hat{n}+1/2)$.
The term depending on $\xi$ is the pseudo-spin Zeeman energy,
and actually coincides with $-m_c \cdot B$.
In graphene, the pseudo-spin Zeeman splitting 
is equal with the Landau level spacing,
so that the $n$-th Landau level at the valley $K_+$
has the same energy $(n+1)$-th level at $K_-$.

\begin{figure*}
\begin{center}
 \leavevmode\includegraphics[height=0.45\hsize]{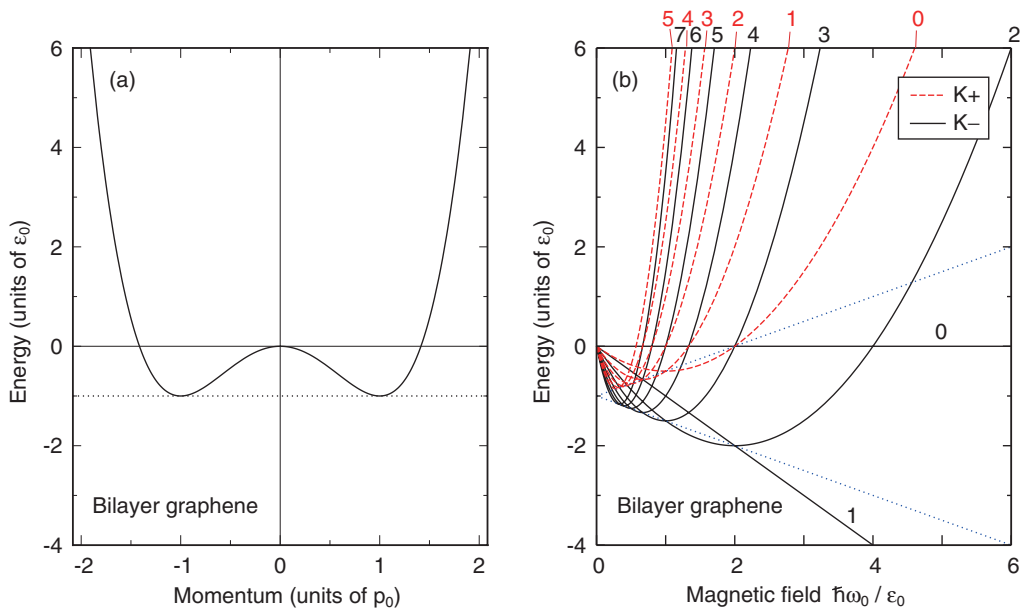}
\end{center}
\caption{
(color online)(a) Low energy dispersion of gapped bilayer graphene
given by Eq.\ (\ref{eq_H_bi_red}).
(b) Landau level spectrum of
Eq.\ (\ref{eq_H_bi_red_B}) with some small $n$'s,
plotted against magnetic field.
Dashed (red) and solid (black) lines represent
the valley $\xi = +$ and $-$, respectively.
Numbers assigned to the curves indicate Landau level index $n$.
A pair of dotted slopes represent 
the energy of the band bottom shifted by pseudo-spin Zeeman energy,
i.e., $-\vare_0 \pm \hbar \omega_0/2$.
At $\Delta = 0.1$ eV, the characteristic energy scale is
 $\vare_0 = 13$ meV and 
the magnetic field for $\hbar\omega_0/\vare_0 = 1$
is 7.6T.
}
\label{fig_spec}
\end{figure*}


The two terms in the current distribution of Eq.\ (\ref{eq_j_mono})
can be distinguished by change in 
the two-dimensional mirror reflection,
\begin{eqnarray}
 F (\Vec{r}) \to F'(\Vec{r}) \equiv F(\Vec{r}'),
\label{eq_reflect}
\end{eqnarray}
where $\Vec{r}=(x,y)$ and $\Vec{r'}=(-x,y)$.
Let $\Vec{j}(\Vec{r})$ and $\Vec{j}'(\Vec{r})$ be 
the expectation values of
the current density for the wavefunctions $F$ and
$F'$, respectively.
Each current component changes with either of $s=\pm$ in
\begin{eqnarray}
  \begin{pmatrix}
   j'_x(\Vec{r}') \\
   j'_y(\Vec{r}')
  \end{pmatrix}
=
s 
  \begin{pmatrix}
   -j_x(\Vec{r}) \\
   j_y(\Vec{r})
  \end{pmatrix},
\label{eq_chiral1}
\end{eqnarray}
or equivalently,
\begin{eqnarray}
  \Vec{r'}\times\Vec{j'}(\Vec{r'})
=
- s \, \Vec{r}\times\Vec{j}(\Vec{r}).
\label{eq_chiral2}
\end{eqnarray}
In Eq.\ (\ref{eq_j_mono}), the first term $\Vec{j}^{\rm (eff)}$
yields to $s=+$, i.e., the current 
map is just mirror-reflected in the same way as $\Vec{r}$.
This is a natural consequence,
since ${\cal H}^{\rm (eff)}$ is invariant in the mirror reflection.

The second term $\Vec{j}_c$ has an opposite sign $s=-$, 
or $\Vec{j}'_c$ goes against
the mirror reflection of $\Vec{j}_c$,
and can be called chiral in this sense.
In gapped graphene, having this term may look counter-intuitive
since the system is originally mirror symmetric
with respect to a line containing an $AB$ bond.
But this ``real'' reflection
exchanges valleys $\xi = \pm$ at the same time
in addition to  Eq.\ (\ref{eq_reflect}),
so that $\Vec{j}_c$ is then simply mirror-reflected 
as it should. Therefore the chiral term 
is necessarily accompanied by the factor $\xi$.

Two current components behave also differently in the 
effective time reversal operation $F\to F^*$
within single valley.
The first term obviously reverses in this operation,
as a consequence of the effective time-reversal symmetry
for ${\cal H}^{\rm (eff)}$.
The second term depends only on the absolute value of the wave amplitude
and thus remains unchanged in the same operation.
But it reverses in the real time-reversal operation
which switches $\xi = \pm$.
We will see that the same argument applies to
bilayer graphene as well.

\section{Bilayer graphene}
\label{sec_bi}
%
Bilayer graphene\cite{Novoselov_et_al_2006a,Ohta_et_al_2006a,Castro_et_al_2007b,Oostinga_et_al_2008a}
is a pair of graphene layers arranged in AB (Bernal) stacking and
includes $A_1$ and $B_1$ atoms on layer 1 and $A_2$ and $B_2$ on layer
2.
\cite{McCann_and_Falko_2006a,Guinea_et_al_2006a,Lu_et_al_2006a,Lu_et_al_2006b,McCann_2006a,Koshino_and_Ando_2006a,Nilsson_et_al_2006b,Partoens_and_Peeters_2006a}
The states at $B_1$ and $A_2$ are coupled by $\gamma_1 \approx 0.39$ eV.
\cite{Misu_et_al_1979}
The low-energy states are described by the Hamiltonian matrix for the
basis $(|A_1\rangle,|B_1\rangle,|A_2\rangle,|B_2\rangle)$,
\cite{McCann_and_Falko_2006a,Guinea_et_al_2006a}
\begin{eqnarray}
{\mathcal H}({\Vec{p}}) =
\begin{pmatrix} 
\Delta & v p_-  & 0 & 0
\\
v p_+ & \Delta & \gamma_1 & 0
\\
0 & \gamma_1  & -\Delta & v p_-
\\
0 & 0   & v p_+ & -\Delta
\end{pmatrix},
\label{eq_H_bi}
\end{eqnarray}
where 
$\Delta$ describes potential asymmetry between 
layer 1 and 2 (not $A$ and $B$ sites), 
which gives rise to
an energy gap. \cite{McCann_and_Falko_2006a,Lu_et_al_2006a,Lu_et_al_2006b,Guinea_et_al_2006a,McCann_2006a,Nilsson_et_al_2006b,Ando_and_Koshino_2009a,Ando_and_Koshino_2009b}
Experimentally the potential asymmetry can be induced 
by applying an electric field perpendicular to the layer,
 \cite{Ohta_et_al_2006a,Castro_et_al_2007b,Oostinga_et_al_2008a,Y_Zhang_et_al_2008, Mak_et_al_2009}
and the asymmetry as large as $\Delta \sim 0.1$ eV was actually observed
in spectroscopic measurements. \cite{Ohta_et_al_2006a,
 Y_Zhang_et_al_2008, Mak_et_al_2009}
For simplicity, we neglected the trigonal warping effect
due to the extra band parameter. \cite{McCann_and_Falko_2006a,Koshino_and_Ando_2006a}

Let us assume $\Delta > 0$ in the following.
At $\Vec{p}=0$, the Hamiltonian gives four eigen energies
\begin{eqnarray}
&& \vare^0_1 = -\sqrt{\gamma_1^2 + \Delta^2}, \quad
 \vare^0_2 = -\Delta,
\nonumber\\
&& \vare^0_3 = \Delta, \quad
 \vare^0_4 = \sqrt{\gamma_1^2 + \Delta^2}.
\label{eq_basis}
\end{eqnarray}
We consider a state near the conduction band bottom 
$\vare = \vare^0_3$, of which wave amplitude is mostly
concentrated on the first component $F_{A1} \equiv F$.
The effective Hamiltonian for $F$ is \cite{Guinea_et_al_2006a}
\begin{eqnarray}
 {\cal H}^{\rm (eff)}({\Vec{p}}) &\approx&
\frac{1}{2\Delta}\frac{v^4p^4}{\gamma_1^2}
- 2\Delta \frac{v^2p^2}{\gamma_1^2}
\nonumber\\
&\equiv&
\frac{{p}^4}{4m_0 p_0^2} - \frac{{p}^2}{2m_0}\
\label{eq_H_bi_red}
\end{eqnarray}
where the energy is measured from $\vare = \Delta$ and 
\begin{eqnarray}
 m_0 = \frac{\gamma_1^2}{4v^2\Delta},
\quad p_0 = \hbar k_0 = \frac{\sqrt{2}\Delta}{v}.
\end{eqnarray}
The term with ${p}^2$ comes from the off-diagonal elements 
${\cal H}_{34}$ and ${\cal H}_{31}$
in the Hamiltonian matrix diagonalized for $p=0$.
To have ${p}^4$ term, 
we need in Eq.\ (\ref{eq_H_red}) the higher order term 
for the off-diagonal matrix element between $j=2$ and 3;
i.e.,use instead of ${\cal H}_{32}$ 
\begin{eqnarray}
 \tilde{\cal H}_{32} = 
{\cal H}_{32} + 
\sum_{m=1,4} {\cal H}_{3n}\frac{1}{\vare^0_3-\vare^0_n}{\cal H}_{n2}.
\end{eqnarray}
The dispersion is plotted in Fig.\ \ref{fig_spec}(a).
It is non-monotonic function of $p$, and
the band minimum appears at off-center momentum $p=p_0$
and energy $\vare = -\vare_0$, where 
\begin{equation}
 \vare_0 = \frac{2\Delta^3}{\gamma_1^2}.
\end{equation}
For instance, the asymmetric energy of $\Delta = 0.1$ eV
gives $\vare_0 = 13$ meV.
The density of states is given by
\begin{equation}
D(\vare) = g_sg_v
\frac{m_0}{2\pi\hbar^2}
\frac{1}{\sqrt{1+\vare/\vare_0}}\times
\left\{
\begin{array}{cl}
0
& (\vare < -\vare_0)
\\
2 & (-\vare_0 < \vare < 0)
\\
1 & (\vare > 0),
\label{eq_dos_bi}
\end{array}
\right.
\end{equation}
where $g_s=g_v=2$ is spin and valley degeneracies.

The local current density of Eq.\ (\ref{eq_j}) is written 
in the same level of approximation as
\begin{eqnarray}
 \langle \Vec{j}(\Vec{r}) \rangle
&=&
{\rm Im}\,\Vec{u}
+ \xi 
(-\Vec{e}_z \times {\rm Re}\,\Vec{u})
\label{eq_j_bi}
\end{eqnarray}
where the vector $\Vec{u}$ is 
defined by 
\begin{eqnarray}
 u_\mu &=& 
-\frac{e\hbar}{2m_0}\frac{1}{k_0^2}
\sum_{\nu=x,y}
\left[
2(\partial_\nu F^*)\partial_\mu(\partial_\nu F)
- \partial_\mu (F^* \partial_\nu^2 F)
\right]
\nonumber\\
&& +\frac{e\hbar}{m_0}F^*\partial_\mu F.
\label{eq_u}
\end{eqnarray}
The second components of $\langle \Vec{j}(\Vec{r}) \rangle$ 
is the chiral current
and expressed as
\begin{eqnarray}
&&  \langle \Vec{j}_c(\Vec{r}) \rangle
=  c\, \nabla \times \GVec{\mu}(\Vec{r}),
\nonumber\\
&& \GVec{\mu}(\Vec{r}) = 
 \xi \Vec{e}_z
\frac{e\hbar}{2 m_0c}
\Bigl\{
- \frac{1}{k_0^2}\
\Bigl[|\nabla F|^2 - {\rm Re}(F^* \nabla^2 F)\Bigr]
+  |F|^2
\Bigr\}.
\nonumber\\
\label{eq_j_bi_chiral}
\end{eqnarray}
The equivalent magnetic moment $\GVec{\mu}(\Vec{r})$ now depends on
$F$ and its derivative.
The magnetization of Eq.\ (\ref{eq_m}) becomes
\begin{eqnarray}
m
&=&
-\frac{e}{2c}
(
x v^{y{\rm (eff)}}
-y v^{x{\rm (eff)}}
)
- \xi 
\frac{e\hbar}{m_0 c}
\left(
\frac{{p}^2}{p_0^2} - \frac{1}{2}
\right).
\nonumber\\
\label{eq_m_bi}
\end{eqnarray}
The second term, $m_c$, is the valley magnetic moment
induced by the chiral current.
The valley splitting energy at the band bottom
can be estimated by inserting $p=p_0$,
\begin{equation}
 2 |m_c(p_0)|\, B = \frac{\hbar e B}{m_0 c}
\equiv \hbar\omega_0.
\end{equation}
The effective $g$-factor for this pseudo-spin splitting
is given by
$g^* = 2m/m_0$ where $m$ is the bare electron mass.
$g^*$ is proportional to $\Delta$
and it approximates $30$ at $\Delta = 0.1$eV.

When the valley splitting exceeds $\vare_F$,
the system is fully valley-polarized 
with single kind of chiral particles.
Using the density of states of Eq.\ (\ref{eq_dos_bi}),
the condition for full valley polarization
is estimated in low $B$-field limit,
\begin{equation}
n < n_{\rm crit} = 
g_s\frac{1}{\pi}\frac{\Delta}{\hbar v}\sqrt{\frac{2eB}{c\hbar}},
\end{equation}
where $n$ is the electron density.
We have $n_{\rm crit} \approx 5\times 10^{11}$ cm$^{-2}$
at $\Delta = 0.1$eV and $B=1$T.
For the gapped monolayer graphene, the condition is
\begin{equation}
 n < n_{\rm crit} = g_s \frac{eB}{h},
\end{equation}
which is approximately $5\times 10^{10}$ cm$^{-2}$ at $B=1$T.
In bilayer, the critical density is proportional to $\sqrt{B}$ rather 
than $B$, and thus the valley polarization
is achieved in much lower magnetic fields
than in monolayer, in a small electron density.
This property is owing to
the divergence of the density of states at the band bottom.

Similarly to monolayer ,
the valley splitting of Landau levels in asymmetric bilayer graphene
\cite{McCann_and_Falko_2006a,Castro_et_al_2007b,Koshino_and_McCann_2009c}
is correctly given by
the pseudo-spin Zeeman energy due to the magnetic moment $m_c$.
The original Hamiltonian in a magnetic field
is given by Eq.\ (\ref{eq_H_bi}) 
with $\Vec{p}$ replaced by ${\GVec{\pi}}$.
Near $\vare = \Delta$, it is reduced to
\begin{eqnarray}
&& {\cal H}^{\rm (eff)}({\GVec{\pi}}) 
 \approx 
\frac{1}{2\Delta} \frac{(v\pi_-)^2(v\pi_+)^2}{\gamma_1^4}
- 2\Delta \frac{(v\pi_-)(v\pi_+)}{\gamma_1^2}
\nonumber\\
&& 
=
\frac{(\hbar\omega_0)^2}{4\vare_0}
\left[
\left(
\hat{n}+\frac{1}{2}+\xi
\right)^2 - \frac{1}{4}
\right]
- \hbar \omega_0 
\left(
\hat{n}+\frac{1}{2}+\frac{\xi}{2}
\right),
\nonumber\\
\label{eq_H_bi_red_B}
\end{eqnarray}
where $\omega_0 = eB/(m_0 c)$.
The pseudo-spin Zeeman energy, i.e., half of the energy difference
between $\xi = \pm$, is transformed to
\begin{eqnarray}
 E_{\rm Zeeman} 
&=&
 \xi 
\frac{e\hbar}{m_0c}
\left(
\frac{{\pi}^2}{p_0^2} 
- \frac{1}{2}
\right) B,
\end{eqnarray}
which coincides with $-m_c \cdot B$ in the limit of $B=0$.

The first and second terms in Eq.\ (\ref{eq_H_bi_red_B})
correspond to ${p}^4$ and ${p}^2$ terms in the 
zero-field Hamiltonian, respectively, 
and become dominant
when $\hbar\omega_0 (n+1/2) \gg \vare_0$
and $\ll \vare_0$, respectively.
In the lower Landau levels where the second term dominates, 
the $n$-th level at the valley $K_+$
and $(n+1)$-th level at $K_-$ approximately degenerate.
In higher levels where the first term becomes dominant,
the $n$-th Landau level at the valley $K_+$
and $(n+2)$-th level of $K_-$ degenerate.
Fig.\ \ref{fig_spec} (b) plots the Landau level energy
of Eq.\ (\ref{eq_H_bi_red_B}) as a function of magnetic field,
where dashed and solid lines represent
the valley $\xi = +$ and $-$, respectively.
At $\Delta = 0.1$ eV, for instance, 
the characteristic the magnetic field corresponding to
$\hbar\omega_0/\vare_0 = 1$ is 7.6T.
A pair of dotted slopes represent 
the energy of the band bottom shifted by pseudo-spin Zeeman energy,
i.e., $-\vare_0 + \xi \hbar \omega_0/2$.
In small $B$-field, they actually serve as the envelope curves for 
Landau levels of $\xi = \pm$.
Full valley polarization occurs below the upper slope.

\begin{figure*}
\begin{center}
 \leavevmode\includegraphics[height=0.4\hsize]{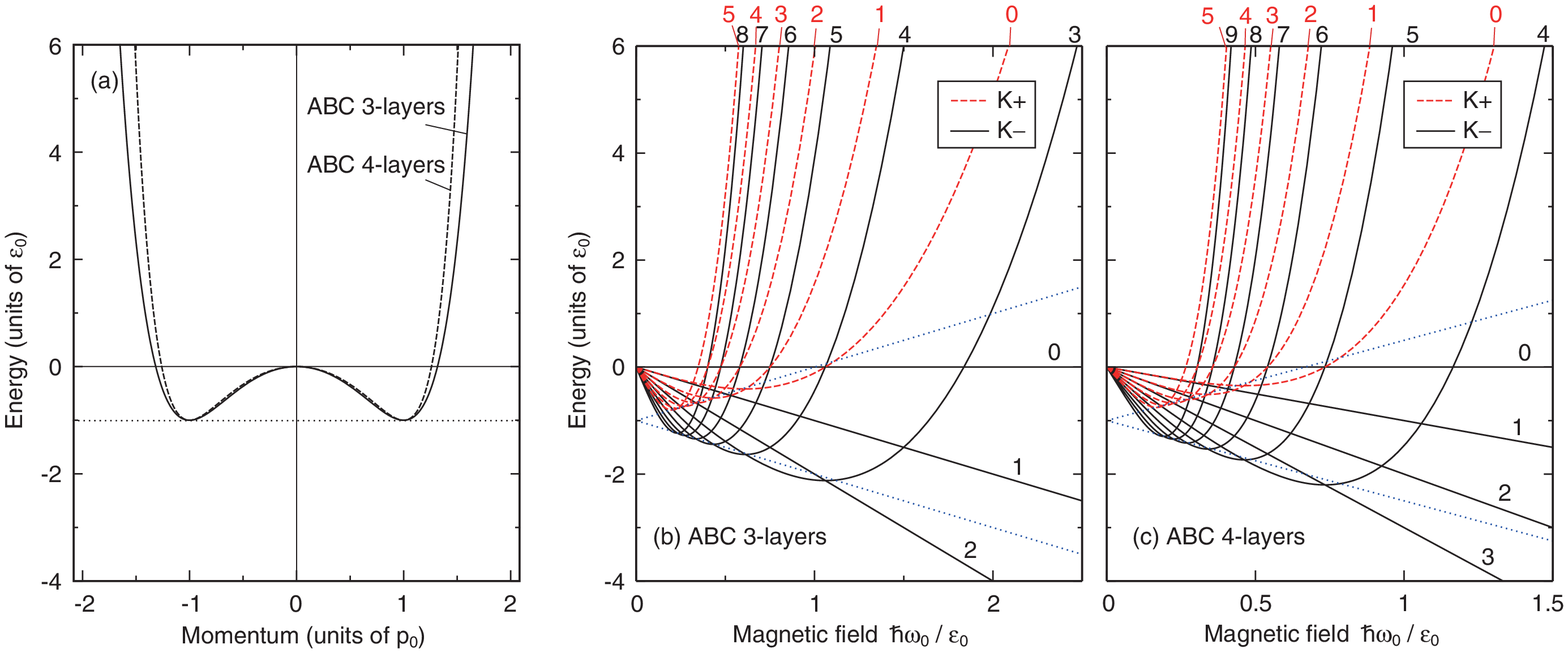}
\end{center}
\caption{
(color online)(a) Low energy dispersion of gapped 
3-layer and 4-layer ABC graphene given by Eq.\ (\ref{eq_H_abc_red}).
(b)(c) Corresponding Landau level spectrum of
Eq.\ (\ref{eq_H_abc_red_B}) with some small $n$'s,
plotted against magnetic field.
A pair of dotted slopes represent 
$-\vare_0 \pm \hbar (N-1)\omega_0/2$.
}
\label{fig_spec_abc}
\end{figure*}

\section{ABC multilayer graphenes}
\label{sec_abc}
%

%

For the structure of bulk graphite, there are two known forms called
ABA (AB, hexagonal, or Bernal) and ABC (rhombohedral) with different
stacking manners.\cite{Lipson_et_al_1942,Haering_1958,McClure_1969}
The ABA phase is thermodynamically stable and common,
while it is known that some portion of natural graphite
takes the ABC form. \cite{Lipson_et_al_1942}
The low-energy band structure of a finite ABC graphene multilayer
is given by a pair for the surface bands localized 
at outer-most layers, \cite{Guinea_et_al_2006a,Manes_et_al_2007,Koshino_2010}
and the interlayer potential asymmetry opens an energy gap between 
those bands. 
\cite{Aoki_and_Amawashi_2007, Lu_et_al_2007,Koshino_and_McCann_2009b,Koshino_2010}

Now we attempt to argue the chiral magnetic moment
of gapped low-energy bands of ABC $N$-layered graphene,
in a parallel way to the bilayer graphene.
If the basis is taken as $|A_1\rangle,|B_1\rangle$;
$|A_2\rangle,|B_2\rangle$; $\cdots$; $|A_N\rangle,|B_N\rangle$,
the low-energy effective Hamiltonian 
can be written as \cite{Guinea_et_al_2006a,Manes_et_al_2007,Lu_et_al_2007,Koshino_and_McCann_2009b,Koshino_2010}
\begin{eqnarray}
 {\cal H}_{\rm ABC} =
\begin{pmatrix}
 H_1 & V  \\
 V^{\dagger} & H_2 & V  \\
 & V^{\dagger} & H_3 & V \\
  &   & \ddots & \ddots & \ddots & 
\end{pmatrix},
\label{eq_H_abc}
\end{eqnarray}
and
\begin{eqnarray}
&& H_j = 
\begin{pmatrix}
 U_j & v p_- \\ v p_+ & U_j
\end{pmatrix}, 
\quad
V = 
\begin{pmatrix}
0 & 0 \\ \gamma_1 & 0
\end{pmatrix},
\label{eq_H_suppl} 
\end{eqnarray}
where 
$U_j$ is the electrostatic potential at $j$th layer.
For simplicity, we neglected the trigonal warping effect
due to the extra band parameter. 
\cite{Koshino_and_McCann_2009b}

The potential asymmetry $U_j$ can be induced 
by applying an electric field ${\cal E}$ perpendicular to the layer.
When ${\cal E}$ is uniform, the potential energy 
with respect to the middle of the stack is
written as
\begin{equation}
 U_j = \left(\frac{N+1}{2} - j \right) e{\cal E}d,
\end{equation}
where $d \approx 0.334$ nm is the interlayer spacing.
The bilayer graphene of Eq.\ (\ref{eq_H_bi}) is  
a special case of Eq.\ (\ref{eq_H_abc})
with $N=2$ and $e{\cal E}d = 2\Delta$.
The actual field ${\cal E}$ 
can be smaller than externally applied electric field
due to the screening by the electrons in the graphene. \cite{Koshino_2010}
We assume ${\cal E} > 0$ and $|U_j| \ll \gamma_1$
in the following.

At $p=0$, there are two low-energy eigenenergies 
at $\vare = U_1$ and $U_N$ 
originating from $|A_1\rangle$ and $|B_N\rangle$,
while all other states appear near $\vare = \pm \gamma_1$
through the dimerization between 
$|B_j\rangle$ and $|A_{j+1}\rangle$ for each of $j = 1,\cdots, N-1$.
The effective Hamiltonian 
for the states near $\vare = U_1$, 
is derived as 
\begin{eqnarray}
 {\cal H}^{\rm (eff)}({\Vec{p}}) &\approx&
\frac{\gamma_1^2}{(N-1)e{\cal E}d}
\left(\frac{vp}{\gamma_1}\right)^{2N}
- e{\cal E}d \left(\frac{vp}{\gamma_1}\right)^{2}
\nonumber\\
&\equiv&
\frac{1}{N}
\frac{p_0^2}{2m_0} 
\left(\frac{p}{p_0}\right)^{2N}
- \frac{{p}^2}{2m_0},
\label{eq_H_abc_red}
\end{eqnarray}
where the energy is measured from $\vare = U_1$ and 
\begin{eqnarray}
 m_0 = \frac{\gamma_1^2}{2v^2(e{\cal E}d)},
\quad p_0 = \frac{\gamma_1}{v}
\left(\sqrt{\frac{N-1}{N}} \frac{e{\cal E}d}{\gamma_1}\right)^{\frac{1}{N-1}}.
\label{eq_m0_abc}
\end{eqnarray}

The term with ${p}^2$ comes from the direct coupling with 
the neighboring dimers formed by $|B_1\rangle$ and $|A_{2}\rangle$,
and ${p}^{2N}$ term is from $N$-th order coupling with
the other low-energy state of $|B_{N}\rangle$.
All other terms are neglected in low energies
as long as $vp_0/\gamma_1 \ll 1$.
The band minimum appears at $p=p_0$
and energy $\vare = -\vare_0$, where 
\begin{equation}
 \vare_0 = \frac{N-1}{N}\frac{p_0^2}{2m_0}.
\label{eq_e0_abc}
\end{equation}
The density of states diverges at $\vare = -\vare_0$ as,
\begin{equation}
D(\vare) \approx g_sg_v
\frac{m_0}{\pi\hbar^2}
\frac{1}{\sqrt{2N}}
\frac{N}{N-1}
\frac{1}{\sqrt{1+\vare/\vare_0}}.
\label{eq_dos_abc}
\end{equation}
For example, we show the energy dispersion of $N=3$ and 4
in Fig.\ \ref{fig_spec_abc}(a).
Note that the unit $p_0$ and $\vare_0$ depend on $N$.
At $e{\cal E}d = 0.2$ eV, for instance,
the characteristic energy scale is $\vare_0 = 54$ meV
and $86$ meV for $N=3$ and 4, respectively. 

The magnetization of Eq.\ (\ref{eq_m}) becomes
\begin{eqnarray}
m
&=&
-\frac{e}{2c}
(
x v^{y{\rm (eff)}}
-y v^{x{\rm (eff)}}
)
\nonumber\\
&&
- \xi 
\frac{e\hbar}{m_0 c}
\left[
\frac{N}{2}
\left(\frac{p}{p_0}\right)^{2(N-1)} 
- \frac{1}{2}
\right],
\label{eq_m_abc}
\end{eqnarray}
where the second term, $m_c$, is the valley magnetic moment.
The valley splitting energy at the band bottom
can be estimated by inserting $p=p_0$,
\begin{equation}
 2 |m_c(p_0)|\, B = (N-1)\hbar\omega_0,
\end{equation}
where $\omega_0 = eB/(m_0 c)$.
The splitting is greater for larger $N$
under the same electric field ${\cal E}$.
The condition for full valley polarization
in low $B$-field limit is
\begin{equation}
n < n_{\rm crit} = 
g_s\frac{1}{\pi}\sqrt{\frac{eB}{c\hbar}}
\frac{\gamma_1}{\hbar v}
\left(
\sqrt{\frac{N-1}{N}}\frac{e{\cal E}d}{\gamma_1}
\right)^\frac{1}{N-1}.
\end{equation}
In the small field region $e{\cal E}d \ll \gamma_1$, which is currently assumed, 
$n_{\rm crit}$ increases for larger $N$,
i.e., the valley polarization is achieved 
up to higher electron density in larger stack.
In the large $N$ limit, $n_{\rm crit}$ approaches
a value independent of ${\cal E}$,
\begin{equation}
n^\infty_{\rm crit} =
g_s\frac{1}{\pi}\sqrt{\frac{eB}{c\hbar}}
\frac{\gamma_1}{\hbar v},
\end{equation}
which approximates $1.5\times 10^{12}$ cm$^{-2}$ at $B=1$T.

The low-energy Landau level spectrum near $\vare = U_1$ is
\begin{eqnarray}
&&{\cal H}^{\rm (eff)}
\approx
\frac{\gamma_1^2}{(N-1)e{\cal E}d}
\frac{(v\pi_-)^N(v\pi_+)^N}{\gamma_1^{2N}}
- e{\cal E}d 
\frac{v\pi_-}{\gamma_1}
\frac{v\pi_+}{\gamma_1}
\nonumber\\
&& \quad
= 
\frac{(N-1)^{N-1}}{N^N}
\frac{(\hbar\omega_0)^N}{\vare_0^{N-1}}
\prod_{j=1}^{N}
\left[
\hat{n}+ j-\frac{1-\xi}{2}N
\right]
\nonumber\\
&&
\qquad\qquad\qquad\qquad
- \hbar \omega_0 
\left(
\hat{n}+\frac{1}{2}+\frac{\xi}{2}
\right).
\label{eq_H_abc_red_B}
\end{eqnarray}
The valley splitting in the limit of $B=0$
is again shown to be equivalent with $-m_c \cdot B$ 
of Eq.\ (\ref{eq_m_abc}).
In higher Landau levels where the first term becomes dominant,
the $n$-th Landau level at the valley $K_+$
and $(n+N)$-th level of $K_-$ degenerate.
Fig.\ \ref{fig_spec_abc} (b) and (c) plot the Landau level spectra
of Eq.\ (\ref{eq_H_abc_red_B}) 
for the cases of $N=3$ and 4, respectively.
The Landau levels in small magnetic fields
are well bound by dotted lines, or
the energies of $-\vare_0 + \xi \hbar \omega_0/2$.
At $e{\cal E}d = 0.2$ eV, for instance, 
the magnetic field corresponding to
$\hbar\omega_0/\vare_0 = 1$ is 33T and 52T for $N=3$ and 4,
respectively.
As argued above, we can see that, 
for greater $N$, the full valley polarization is possible
up to larger electron density (i.e., more Landau levels)
at the same magnetic field.


\section{Pseudo-spin paramagnetism}
\label{sec_pseudo}

The pseudo-spin Zeeman splitting causes the Pauli paramagnetism 
in analogous way to real spin.
The magnetic susceptibility was previously calculated
for gapped monolayer and bilayer graphenes \cite{Koshino_and_Ando_2010}, 
and it was shown that the susceptibility 
in the quadratic dispersion near $K_\pm$ point,
is expressed as sum of valley pseudo-spin paramagnetism 
and Landau diamagnetism similarly to a bare electron.
In monolayer graphene, the pseudo-spin paramagnetism 
diverges in the zero gap limit, 
leading to a singular orbital susceptibility
where the strong diamagnetism suddenly disappears off the Dirac point.
 \cite{McClure_1956a,Sharapov_et_al_2004a,Fukuyama_2007a,Nakamura_2007a,Koshino_and_Ando_2007b,Ghosal_et_al_2007a,Koshino_et_al_2009a,Koshino_and_Ando_2010}

Here we extend the argument to general electronic structures
other than quadratic,
and show that the pseudo-spin splitting always accompanies
paramagnetic contribution in any part of the dispersion.
Let us consider a system in a magnetic field $B$
with the Landau level sequence,
\begin{eqnarray}
&& \vare_{n} = \vare (x_n, \delta) 
\quad (n=0,1,2,\cdots),
\label{eq_en}
\nonumber\\
&& x_n = \left(n+\frac 1 2 \right)\delta, \quad
\delta = \hbar\omega_c = \frac{\hbar e B}{m^*c},
\label{eq_x}
\end{eqnarray}
where $n$ is the Landau level index,
$m^*$ is the effective mass characterizing the system.
The second argument $\delta$ in $\vare(x_n, \delta)$
represents the dependence on $B$, 
which are not included in $x_n$.
For example 
the low-energy Landau level of gapped monolayer graphene,
Eq.\ (\ref{eq_H_mono_red_B}),
is given by
\begin{eqnarray}
 \vare(x_n,\delta) = x_n + \frac{\xi}{2} \delta,
\label{eq_e_mono}
\end{eqnarray}
and that of bilayer graphene,
Eq.\ (\ref{eq_H_bi_red_B}), by
\begin{eqnarray}
 \vare(x_n,\delta) = 
\frac{1}{4\vare_0}
\left[(x_n+\xi\delta)^2 - \frac{1}{4}\delta^2\right] + 
\left(x_n + \frac{\xi\delta}{2}\right),
\label{eq_e_bi}
\end{eqnarray}
with $m^*$ replaced by $m_0$.

By treating $x(=x_n)$ and $\delta$ as independent variables,
we can expand $\vare(x,\delta)$ as 
\begin{equation}
\varepsilon (x, \delta) = \varepsilon^{(0)}(x)
+ \varepsilon^{(1)}(x)\delta + \frac{1}{2}\varepsilon^{(2)}(x)\delta^2 + \cdots.
\label{eq_expand}
\end{equation}
The zero-th order term $\vare^{(0)}$ 
is related to the energy spectrum at $B=0$.
When the system is isotropic, in particular,
the dispersion is given by $\vare^{(0)} (x)$ with $x = p^2/2m^*$.
The first order shift $\vare^{(1)} \delta$
can be regarded as pseudo-spin Zeeman term
associated with magnetic moment $-(e\hbar/cm^*) \,\vare^{(1)}$,
which corresponds to $m_c$ in previous arguments.

The thermodynamic potential becomes
\begin{eqnarray}
&& \Omega = -\frac{1}{\beta} \frac{1}{2\pi l_B^2}
\sum_{n=0}^\infty 
\varphi\big[\varepsilon(x_n,\delta) \big]
\nonumber\\
&& =
 -\frac{1}{\beta} \frac{m^*}{2\pi\hbar^2}
\Bigg[ \! \int_{0}^\infty \!\!
\varphi\big[\varepsilon(x,\delta) \big]  dx 
 + \frac{\delta^2}{24} 
\frac{\partial \varphi[\varepsilon(x,0)]}{\partial x}
\Big|_{x=0}
 \Bigg] 
\nonumber\\
&&
\qquad + O(\delta^3),
\label{eq:omega_bilayer}
\end{eqnarray}
where $\varphi(\vare) = \ln[1+e^{-\beta(\vare-\mu)}]$,
$\beta = 1/(k_B T)$, $\mu$ is the chemical potential,
and we used the Euler-Maclaurin formula in the second equation.
Using Eq. (\ref{eq_expand}), we can further expand
$\Omega$ in terms of $\delta \propto B$.
The magnetization is given by
\begin{equation}
 M = - \left(\frac{\partial \Omega}{\partial B}\right)_\mu,
\end{equation}
and the magnetic susceptibility by
\begin{equation}
\chi 
= -\Big(\frac{\partial^2 \Omega}{\partial B^2}\Big)_\mu \Big|_{B=0}.
\label{eq:chi_def}
\end{equation}
We end up with
\begin{eqnarray}
 \chi(\mu,T) = \int_{-\infty}^\infty d\vare
\left(
-\frac{\partial f}{\partial \vare}
\right)
\chi(\vare),
\end{eqnarray}
with
\begin{eqnarray}
\chi(\varepsilon) \! &=& \! 
\left(\frac{e\hbar}{c m^*}\right)^2 
\Biggl[
D(\varepsilon) \bigl(\varepsilon^{(1)}\bigr)^2 
- \int_{}^\varepsilon d\varepsilon D (\varepsilon)  
\varepsilon^{(2)},
\nonumber\\
&& \!\!\! 
- \frac{1}{12} 
\frac{m^*}{2\pi\hbar^2}
\theta\bigl(\varepsilon-\varepsilon^{(0)}(0)\bigr)
\frac{\partial \varepsilon^{(0)}(x)}{\partial x} \Big|_{x=0}
\Biggr]. \quad
\label{eq_chi_form}
\end{eqnarray}
where $f(\vare) = \left[1 + e^{\beta(\vare - \mu)}\right]^{-1}$
is the Fermi distribution function, 
and $\varepsilon^{(1)}$  and $\varepsilon^{(2)}$ 
are regarded as functions of energy $\varepsilon$
through $\varepsilon=\varepsilon^{(0)}(x)$.
$D(\vare)$ is the density of states given by
\begin{equation}
 D (\vare) = \frac{m^*}{2\pi \hbar^2} \int_0^\infty 
\delta(\vare - \vare^{(0)}(x)) dx.
\end{equation}
The susceptibility at $T=0$ is given by $\chi(\mu)$.
The first term in Eq.\ (\ref{eq_chi_form})
is regarded as the Pauli paramagnetism induced by 
the pseudo-spin magnetic moment.
It is always positive,
and purely determined by the density of states 
and the magnetic moment at Fermi energy.
The second term is the summation of the second order energy shift
$\vare^{(2)}$ over all the states below Fermi level,
and the third term gives a discrete jump at 
the energy corresponding to $p=0$.

For the low-energy spectrum of 
the gapped monolayer graphene, Eq.\ (\ref{eq_e_mono}), 
we obtain \cite{Koshino_and_Ando_2010} 
\begin{eqnarray}
&& \chi = \chi_P + \chi_L
\nonumber\\
&&
 \chi_P = D \mu^{*2}_B,
\quad  
 \chi_L = - \frac{1}{3} D \mu^{*2}_B,
\label{eq_chi_mono}
\end{eqnarray}
where $\chi_P$ and $\chi_L$ come from
the first and the third terms
in Eq.\ (\ref{eq_chi_form}), respectively,
and the second term is zero.
Here $D = g_s g_v m/(2\pi\hbar^2) \, \theta(\vare)$
is the density of states,
$\mu^*_B=e\hbar/(2m^*c)$
is the effective Bohr magneton.
Obviously $\chi_P$ and $\chi_L$ correspond to conventional 
Pauli paramagnetism and Landau diamagnetism, respectively.
The susceptibility calculated above is
the contribution from the conduction band,
while the valence band gives the exactly opposite jump
at the valence band top.
The total susceptibility is 
diamagnetic at $\chi = -\chi_P-\chi_L$ in the gap region,
and disappears in conduction and valence bands. \cite{Koshino_and_Ando_2010}

\begin{figure}
\begin{center}
 \leavevmode\includegraphics[width=1.\hsize]{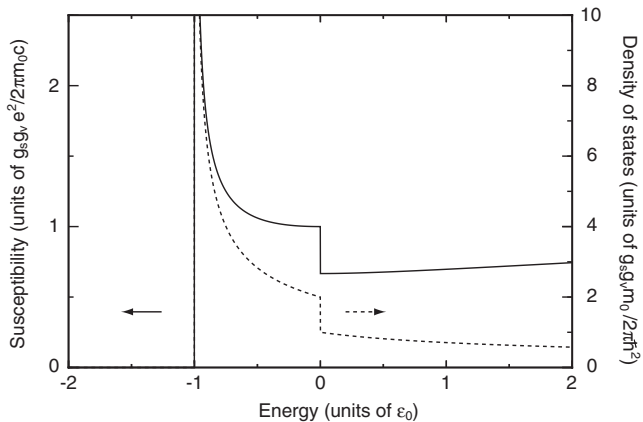}
\end{center}
\caption{Susceptibility (solid) 
and density of states (dashed)
near the band bottom of asymmetric bilayer graphene,
plotted against the Fermi energy. 
}
\label{fig_chi}
\end{figure}

For gapped bilayer graphene, Eq.\ (\ref{eq_e_bi}), we get
\begin{eqnarray}
 \chi(\vare) &=&
\frac{ g_sg_v e^2}{2\pi m_0 c^2}\times
\left\{
\begin{array}{l}
0
\quad (\vare < -\vare_0),
\\
\displaystyle
 \frac{1}{2}
\frac{2+\vare/\vare_0}{\sqrt{1+\vare/\vare_0}}
\quad (-\vare_0 < \vare < 0),
\\
\displaystyle
 \frac{1}{4}
\frac{2+\vare/\vare_0}{\sqrt{1+\vare/\vare_0}}
+ \frac{1}{6}
\quad (\vare > 0).
\end{array}
\right.
\nonumber\\
\label{eq_chi_bi}
\end{eqnarray}
The susceptibility diverges at the band bottom,
$\vare = -\vare_0$. \cite{Koshino_and_Ando_2010}
The physical meaning of the divergence is obvious,
since the Pauli paramagnetism, i.e., the first term
of Eq.\ (\ref{eq_chi_form}) is proportional to
the density of states, which diverges at the band bottom.
The susceptibility of Eq.\ (\ref{eq_chi_bi})
is plotted in Fig.\ \ref{fig_chi}
together with and the density of states, Eq.\ (\ref{eq_dos_bi}).

The argument can be extended to ABC $N$-layer graphene
in a straightforward fashion.
Using Eqs.\ (\ref{eq_dos_abc}) and (\ref{eq_m_abc}),
the pseudo-spin paramagnetic susceptibility 
above and near the band bottom $\vare = -\vare_0$
is written as
\begin{eqnarray}
 \chi(\vare) &\approx&
D(\vare) m_c^2
=
\frac{g_sg_v e^2}{2\pi m_0 c^2}
\frac{N(N-1)}{2\sqrt{2N}}
\frac{1}{\sqrt{1+\vare/\vare_0}},
\nonumber\\
\label{eq_chi_abc}
\end{eqnarray}
where $m_0$ and $\vare_0$ are defined in
Eqs.\ (\ref{eq_m0_abc}) and (\ref{eq_e0_abc}),
respectively.
The paramagnetic divergence is stronger for greater $N$.

\section{Space-dependent Hall conductivity}
\label{sec_hall}

If the system is modulated by an external scalar potential,
the anomalous current term gives a response current 
in analogous way to the Hall effect.
Here we argue the relation of the associated
Hall conductivity to the anomalous magnetic moment.
In graphenes, such the Hall current exactly cancels between two valleys
due to the time-reversal symmetry,
and the real current appears only when the valley populations
are differentiated, like the Pauli paramagnetism. 
In odd-valley case such as 
the surface states of strong topological insulator,
it directly gives a net current and causes
a magneto-electric response.
\cite{Qi_et_al_2008,Essin_et_al_2009} 

We consider a current distribution
in a finite and isolated system 
modulated by an external potential $V(\Vec{r})$.
In the current densities of 
gapped monolayer and bilayer graphenes,
given by Eq.\ (\ref{eq_j_mono}) and Eq.\ (\ref{eq_j_bi}),
respectively, 
the first term cancels in summation over the occupied states
because it reverses the effective time-reversal operation
$F \to F^*$.
Then the total current is given by a summation of
the chiral term $c\nabla\times\GVec{\mu}(\Vec{r})$ as
\begin{eqnarray}
\Vec{J}(\Vec{r}) &=& 
c \nabla \times \Vec{M}(\Vec{r}),
\nonumber\\
\Vec{M}(\Vec{r}) &=& 
\sum_{\rm occupied}  \GVec{\mu}(\Vec{r}).
\label{eq_jtot}
\end{eqnarray}

When the potential $V(\Vec{r})$ 
is weak and slowly-varying, the Thomas-Fermi approximation gives
\begin{eqnarray}
 M(\Vec{r}) \approx M_F - \partd{M_F}{\vare_F}V(\Vec{r}),
\end{eqnarray}
where $M_F$ is the total magnetization of a uniform system,
\begin{eqnarray}
  M_F = \frac{1}{(2\pi\hbar)^2}\int_{\rm occupied} m_c(p) d^2\Vec{p},
\label{eq_mf}
\end{eqnarray}
and 
$m_c(p)$ being the anomalous magnetic moment at the momentum $p$.
Then Eq.\ (\ref{eq_jtot}) becomes
\begin{eqnarray}
\Vec{J}(\Vec{r}) 
&=&
 ce \partd{M_F}{\vare_F} [\Vec{e}_z \times \Vec{E}(\Vec{r})],
\end{eqnarray}
where $\Vec{E}(\Vec{r}) = - \nabla V(\Vec{r})/(-e)$ is
the electric field, leading to
a response function
\begin{equation}
 \sigma_{xy} = -  ce \partd{M_F}{\vare_F}.
\label{eq_hall}
\end{equation}

By applying Eq.\ (\ref{eq_hall}) to the conduction band 
electrons of gapped monolayer graphene, 
where $m_c(p)$ is given by the second term of Eq.\ (\ref{eq_m_mono}),
we have
\begin{equation}
 \sigma_{xy} =\xi \frac{e^2}{2h}.
\label{eq_hall_mono}
\end{equation}
For gapped $N$-layer ABC graphenes including bilayer,
of which $m_c(p)$ is given by the second term of Eq.\ (\ref{eq_m_abc}),
the expression approximates in high energies $\vare \gg \vare_0$,
\begin{equation}
 \sigma_{xy}  \approx \xi\frac{Ne^2}{2h}.
\end{equation}

\begin{figure}
\begin{center}
 \leavevmode\includegraphics[width=0.9\hsize]{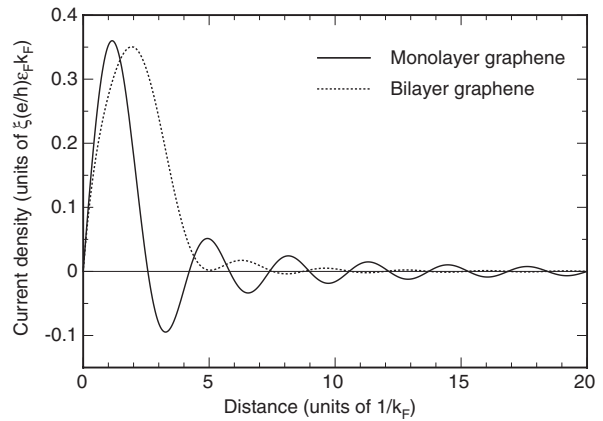}
\end{center}
\caption{
Single-valley current distribution 
contributed by conduction band electrons 
of gapped monolayer (solid) and bilayer graphenes (dashed),
terminated at $x=0$.
}
\label{fig_edge}
\end{figure}

When the system is confined to a finite space,
the above-mentioned Hall current gives 
a chiral edge current at the boundary.
When the confining potential is slowly varying in space,
the current circulation is
\begin{equation}
 I = - \frac{1}{e} \int^{\vare_F} \sigma_{xy}(\vare) d\vare = c M_F,
\label{eq_edge_current}
\end{equation}
as a natural consequence.
This is equally true in a sharp potential as well, where the current 
is distributed in a range of the Fermi wave length from the boundary.
Fig.\ \ref{fig_edge} illustrates the single-valley current distribution
given by the conduction band electrons 
of gapped monolayer and bilayer graphenes.
The detail of the derivation is presented in Appendix.

From its definition, 
the Hall conductivity argued here, Eq.\ (\ref{eq_hall}), 
is the long wavelength limit of the static Hall conductivity,
namely, $\lim_{q\to 0} \lim_{\omega\to 0} \sigma_{xy}(q,\omega)$.
For the original Hamiltonian of monolayer graphene, Eq.\ (\ref{eq_H_mono}),
this is evaluated as \cite{Ludwig_et_al_1994a}
\begin{eqnarray}
 \lim_{q\to 0} 
\lim_{\omega\to 0}
\sigma_{xy}(q,\omega) &=& - \xi \frac{e^2}{2h} \theta(\Delta-|\vare_F|),
\label{eq_hall1}
\end{eqnarray}
where $\theta(x) = 1 \,(x>0), 0 \,(x<0)$ is the step function,
and $\Delta>0$ is assumed here.
The low-energy result, Eq.\ (\ref{eq_hall_mono}),
describes the contribution from the conduction band electrons,
and indeed coincides with 
the discontinuous jump at $\vare = \Delta$ in Eq.\ (\ref{eq_hall1}).
The valence band gives an exactly opposite jump at $\vare=-\Delta$,
so that we have the half-integer Hall conductivity inside the gap,
and zero in the conduction and valance bands.

Note that usual Hall conductivity relevant in the transport
is given by a different limit,
$\lim_{\omega\to 0} \lim_{q\to 0} \sigma_{xy}(q,\omega)$.
This is calculated for gapped monolayer graphene as, \cite{Sinitsyn_et_al_2008}
\begin{eqnarray}
\lim_{\omega\to 0}
 \lim_{q\to 0}  
\sigma_{xy}(q,\omega) &=& 
\left\{
\begin{array}{lc}
\displaystyle
- \xi \frac{e^2}{2h} \frac{\Delta}{|\vare_F|} & (|\vare_F|>\Delta), 
\\
\displaystyle
- \xi \frac{e^2}{2h} & (|\vare_F|<\Delta), 
\end{array}
\right.
\label{eq_hall2}
\end{eqnarray}
which differs from Eq.\ (\ref{eq_hall1}) except for the value inside the gap.
The Berry curvature 
is directly related to this transport Hall conductivity. 
\cite{Thouless_et_al_1982,Xiao_et_al_2010}

From the relationship between the local current and local magnetic
moment, Eq.\ (\ref{eq_jtot}),
the spatial-dependent static Hall conductivity $\sigma_{xy}(q)$ 
can be formulated as a magnetization-density correlation function, i.e.,
\begin{equation}
M(q) = \frac{1}{e} \sigma_{xy}(q) V(q).
\end{equation}
In the low-energy region of gapped monolayer graphene,
it becomes a density-density correlation function,
because the pseudospin magnetization Eq.\ (\ref{eq_m_mono}) is constant 
for each eigenstate regardless of the detail of the wavefunction.
This suggests that $\sigma_{xy}(q)$
is insensitive to the disorder localization effect 
since the magnetic moment of each eigenstate remains
even when the wavefunction is localized. 
This is in contrast to the transport Hall conductivity,
where the localized eigenstates have zero contribution.

Lastly, we show that Hall conductivity Eq.\ (\ref{eq_hall})
is directly related to
index difference $\Delta n$ between degenerated Landau levels of two valleys,
which are argued in the previous sections.
This is defined by the ratio of pseudo-spin Zeeman splitting to
Landau level spacing, or
\begin{eqnarray}
&& \Delta n =\frac{2m_c B}{\hbar\omega_c},
\nonumber\\
&& \hbar\omega_c = \frac{\hbar eB}{c}2\pi
\left(\partd{S(\vare_F)}{\vare_F}\right)^{-1}
\end{eqnarray}
and  $S(\vare_F)=\pi p_F^2$ is the area of the momentum space 
at the Fermi energy $\vare_F$. 
Using Eqs.\ (\ref{eq_mf}) and (\ref{eq_hall}), we obtain,
\begin{eqnarray}
 \Delta n =  \frac{2hc}{e}\partd{M_F}{\vare_F} = 
-\frac{2h}{e^2}\sigma_{xy}.
\end{eqnarray}
Indeed, we have $\Delta n =1$ for gapped monolayer graphene,
and  $\Delta n \approx N$  for gapped $N$-layer ABC graphene 
(including bilayer graphene) in high energies.

\section{Conclusion}
\label{sec_concl}

We presented systematic analyses 
of anomalous chiral current and magnetic moment
in gapped graphenes and related materials.
Starting from the low-energy effective-mass theory,
we formulate a description of local current distribution
supporting anomalous magnetic moment
in general Bloch systems.
In gapped monolayer, bilayer and ABC multilayer graphenes,
we showed that the chiral current circulation accounts for
the valley-dependent magnetic moment
and valley-splitting of Landau levels.
The bilayer and ABC multilayer graphenes exhibit 
a large paramagnetism
at the band bottom, and full valley polarization 
is possible in relatively high electron density.

There have been suggested various mechanisms
for valley polarization or valley filtering
which might be used to control electronic devices.
\cite{Xiao_et_al_2007,Yao_et_al_2008,Rycerz_et_al_2007,Pereira_et_al_2009,Abergel_and_Chakraborty_2009,Nakanishi_et_al_2010}
The possibility of full valley polarization in graphene bilayer
and ABC multilayers
invokes a simple mechanism for valley-dependent transport.
For example, if we could locally apply opposite magnetic fields
to the left and right sides of a gapped bilayer
or ABC-multilayer strip,
and achieve different valley polarizations in two regions,
then the transport between two regions would be killed,
as long as the valley flipping is prohibited in the intermediate region,
i.e., the impurity potential and the spacial magnetic field change 
are smooth compared to the atomic scale.
On the contrary, electrons 
can travel almost 
freely when the same magnetic field is applied to two regions.

While we focus on the family of ABC-stacked multilayer graphenes
in the present studies, the anomalous magnetic moment arises
in ABA-stacked multilayer graphenes as well
when the inversion symmetry is broken. \cite{Koshino_and_McCann_2009c}
In ABA multilayers with an odd number of layers,
the lattice structure originally lacks in the inversion symmetry
so that the valley splitting intrinsically 
exists even in absence of the external field.
\cite{Koshino_and_McCann_2011}
The present analysis applies to every subband comprising the 
total band structure, each of 
which is akin to gapped monolayer or bilayer graphenes. 
\cite{Koshino_and_McCann_2011,Koshino_and_Ando_2009_ssc}

\section*{Acknowledgment}

This project has been funded by JST-EPSRC
Japan-UK Cooperative Programme Grant No. EP/H025804/1.

\appendix

\section{Chiral edge current}
\label{sec_app}

Here we calculate the edge current distribution
of gapped monolayer and bilayer graphenes
bound by a sharp confining potential.
Let us consider a low-energy Hamiltonian
gapped monolayer graphene, Eq.\ (\ref{eq_H_mono_red}),
bound by a potential barrier, 
\begin{equation}
 V(x)
=\left\{
\begin{array}{cl}
 \infty & (x<0)
\\
 0 & (x>0)
\end{array}
\right. .
\label{eq_pot}
\end{equation}
The eigenstates are given by
\begin{eqnarray}
&& F(\Vec{r}) \propto e^{i k_y y} \sin k_x x.
\end{eqnarray}
The current density of Eq.\ (\ref{eq_j_mono})
integrated over the occupied states is written 
in terms of the Bessel function as,
\begin{eqnarray}
J_y(\Vec{r}) &=& 
\xi\frac{e}{h}
\frac{\vare_F}{x}J_2(2k_F x).
\end{eqnarray}
It oscillates and decays in the length scale of $2\pi/k_F$
as shown in Fig.\ \ref{fig_edge}.
The total edge current is
\begin{eqnarray}
 I &\equiv& \int_0^\infty dx\, 
J_y(\Vec{r}) 
= \xi \frac{e}{2h} \vare_F,
\end{eqnarray}
which coincides with $c M_F$.

The similar argument is available in bilayer graphene.
For simplicity, we 
consider high energies $\vare \gg \vare_0$
and neglect ${p}^2$ term in Eq.\ (\ref{eq_H_bi_red}).
The Schr\"{o}dinger equation becomes
the fourth-order differential equation due to the 
${p}^4$ term, and the boundary condition
becomes $F(0)=F'(0)=0$. The eigenstate then becomes
\begin{eqnarray}
&& F(\Vec{r}) \propto e^{i k_y y}
[\cos k_x x + \sin k_x x - e^{-k_x x}].
\end{eqnarray}
The total current density, Eq.\ (\ref{eq_j_bi_chiral}), 
integrated over the occupied states
is numerically calculated and plotted in
Fig.\ \ref{fig_edge}.
The length scale is again characterized by is Fermi wave length,
but it decays more rapidly than in monolayer.
The total edge current is shown to be
\begin{eqnarray}
 I = \int_0^\infty dx\, 
J_y(\Vec{r}) = \xi \frac{e}{h} \vare_F,
\label{eq_i_bi}
\end{eqnarray}
which is twice as large as monolayer's.

%
%
\end{document}